\documentclass[11pt,twoside]{article}
\usepackage{asp2010,graphicx}

\resetcounters

\begin{document}

\title{CIV Emission as a Probe of Accretion Disk Winds}
\author{Gordon T.\ Richards,$^1$
\affil{$^1$Department of Physics, Drexel University, 3141 Chestnut Street, Philadelphia, PA 19104}
}

\begin{abstract}
We present a brief description of a model for the broad emission line region (BELR) in quasars, which is supported by analysis of CIV and other emission lines in the spectra of high-$z$ SDSS quasars.  Specifically we consider a two-component BELR with a disk and wind where the relative strength of each component is a function of luminosity.  The implications of such a model for our understanding of quasar outflows and estimates of their black hole masses and accretion rates are discussed.
\end{abstract}

\section{A Modern Picture of the Broad Emission Line Region}

We argue that emission lines are, in some ways, better diagnostics of AGN winds than absorption lines because they allow us to learn something about winds in every object and not just the fraction that show intrinsic absorption.  As such, we start with a picture of the broad emission line region (BELR) and explain how this picture is relevant to the question of AGN winds

In this model, one source of BELR gas is the (possibly warped) outer accretion disk, which is illuminated by the central engine/inner accretion disk \citep{cdm+88}.  At low luminosity, this component is the dominant source of BEL photons in AGNs.  The atmosphere of the accretion disk in these sources is permeated by an MHD wind \citep[e.g.,][]{kon06}.  In more luminous quasars with relatively softer spectral energy distributions (SEDs) a strong radiation line driven wind forms in the UV emitting part of the accretion disk and dominates the MHD wind \citep{Proga03}.  Such a wind forms when there are enough UV photons to push on the atoms (through line transitions) in the atmosphere of the disk, while, at the same time, there are few enough ionizing photons that those same atoms are not over-ionized \citep{mcgv95,psk00}.  This (radiation line-driven) wind region is the other source of BEL gas and is dominated by high-ionization line transitions.  
Because of filtering through the wind \citep{Leighly04,lc07}, the disk sees a different continuum than the wind (and possibly different from what we see) and the strength of the disk component is inversely related to the strength of the wind component.

In this picture, the fundamental {\em observed} parameter is the SED---nearly all of the diversity in quasar spectra comes from differences in the SED, which, in turn, causes differences in the emission lines due to the changing ratio of the wind to disk component of the BELR.  In short, the harder the SED, the less radiation line driving dominates and the stronger the disk component of the BELR.  Mass and accretion rate are assumed to be degenerate with the SED; the SED serves a surrogate for these more fundamental parameters.  
Indeed it may be that most other parameters are also degenerate.  For example, covering fraction clearly produces differences in {\em individual objects}, but the global covering fraction may be degenerate with the SED in the sense that objects with certain SEDs are going to drive certain winds that yield certain covering fractions.  

By way of summary, in this picture the answer to the question of the origin of the $\sim$20\% BALQSO fraction \citep[e.g.,][]{hf03} is as follows.  Objects with relatively hard SEDs (such as radio-loud quasars and Seyferts) are not able to produce a strong radiation line-driven wind.  The weaker the line driven wind, the less likely BAL outflows are to be present (along {\em any} line of sight).  On the other hand, objects with relatively soft SEDs are the parent sample of BALQSOs (whether we see the troughs along our line of sight or not).  Despite heavy absorption in CIV, the true nature of BALQSOs are revealed by their CIII] {\em emission-line} profiles \citep{rrh+03,Richards2006}.  LoBALs are those objects with the strongest radiation line-driven winds; they are the most luminous and have the bluest optical continua (despite heavy dust reddening). 

Surprisingly, orientation seems to matter relatively little for explaining differences in UV BEL properties.  For example, our claim to have identified the parent sample of BALQSOs is essentially equivalent to saying that orientation must not completely obscure the underlying nature of the BELR in quasars with strong line driven winds \citep{Richards2011}.
Furthermore, we have shown that 1) the (orientation-dependent) {\em morphological} radio properties of quasars have no preference for SED/emission line properties of one type and 2) that it is possible to construct a sample of radio-quiet quasars that are indistinguishable from radio-loud quasars in terms their BELR properties, further supporting the hypothesis that orientation is not a significant cause of diversity in quasar classes (which is not the same as saying that it doesn't matter for individual objects) \citep{Richards2011}.

One might ask why/how quasars might have a range of wind structures and SEDs.  A better question would be why would two quasars with different masses and accretion rates NOT have different SEDs and thus different wind structures (and BEL properties).  Indeed the burden of proof falls squarely upon the static model of quasars.

\section{Evidence for this Model}

Evidence for this model of the BELR and its connection to winds comes from a long history of investigation into the so-called ``Eigenvector 1 (EV1)'' relationships \citep[e.g.,][]{bg92,bf99,smd00} at low-redshift, coupled with two key parameters that can be used to distinguish quasar extrema at high redshift.

The first of those parameters is the so-called Baldwin Effect \citep{Baldwin77}, whereby more luminous quasars are seen to have weaker CIV emission (in term of equivalent width [EQW]).  The second parameter is the blueshifting of the CIV emission line (relative to the expected laboratory wavelength) as first noted nearly 30 years ago \citep{gas82}.   In \citet{rvr+02} and \citet{Richards2011} we have shown that these blueshifts are not anomalies, but rather are nearly ubiquitous in the quasar population.

In Figure~\ref{fig:fig1} we show the joint distribution of CIV EQW and blueshift, thus combining the two parameters above.  We see that the average radio-quiet quasar has a CIV emission line with a rather large blueshift (nearly $1000\,{\rm km\,s^{-1}}$) and that radio-loud quasars span only part of the space that RQ quasars do (in the top-left corner).  On the other hand, BALQSOs (both with troughs observed and expected along other lines of sight) tend towards the lower right-hand corner \citep{Richards2011}.  

\articlefigure[scale=0.5]{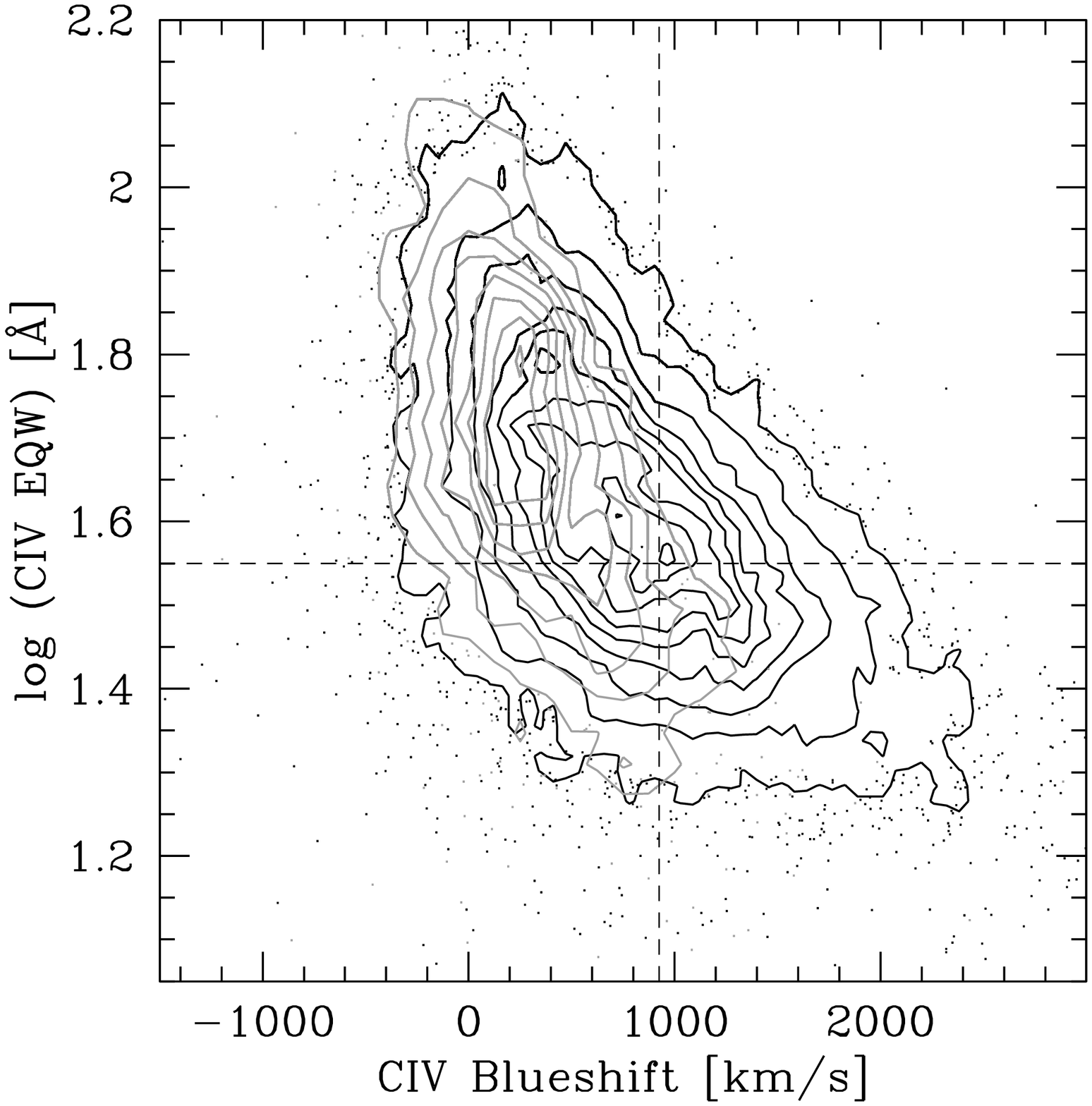}{fig:fig1}{CIV EQW vs.\ CIV blueshift for $\sim$30k SDSS quasars. Radio-quiet (RQ) quasars are shown as black contours/ dots, while radio-loud (RL) quasars are shown as grey contours/dots. The dashed lines indicate the mode of the RQ quasar distribution.  This plot is similar to Figure 7 in \citet{Richards2011}, except that the CIV EQW values have been corrected for intrinsic narrow absorption (Bowler, Allen, \& Hewett, in preparation).  The systemic redshifts for the emission lines were taken from \citet{HW10}.  While RL (and hard-spectrum RQ) quasars are biased towards the top-left in this distribution, soft-spectrum BALQSOs are biased towards the bottom-right.  It is suggested that, while all quasars have outflows of some sort, this diagram enables the identification of those quasars with outflows that are dominated by radiation line driving.}

We have argued \citep{Richards2011,krg+11} that the trend seen in Figure~\ref{fig:fig1} is due to differences in the SED across this dual CIV parameter space.  This is best seen by exploring the full range of emission line differences as a function of composites in the extremes of this distribution as shown in Figure~\ref{fig:fig2}.  The quasars in the top-left of Figure~\ref{fig:fig1} show evidence for a relatively hard SED and a strong disk component of the BELR, while the quasars in the bottom-right of Figure~\ref{fig:fig1} show evidence for a relatively soft SED and a strong wind component of the BELR \citep{krg+11}.  Further supporting evidence comes from connecting this work at high-redshift to EV1 parameters at low-redshift \citep[e.g.,][]{sbm+07}.  Importantly, the unseen EUV continuum may play an important role in determining the relative strength of the wind and disk components of the BELR.  The EUV region is a long-standing problem in understanding the BELR \citep[e.g.,][]{nd79}; our contribution is merely to point out that excess EUV flux is not needed in all objects---just to explain those that have disk-dominated BELRs.




\articlefigure[scale=0.6]{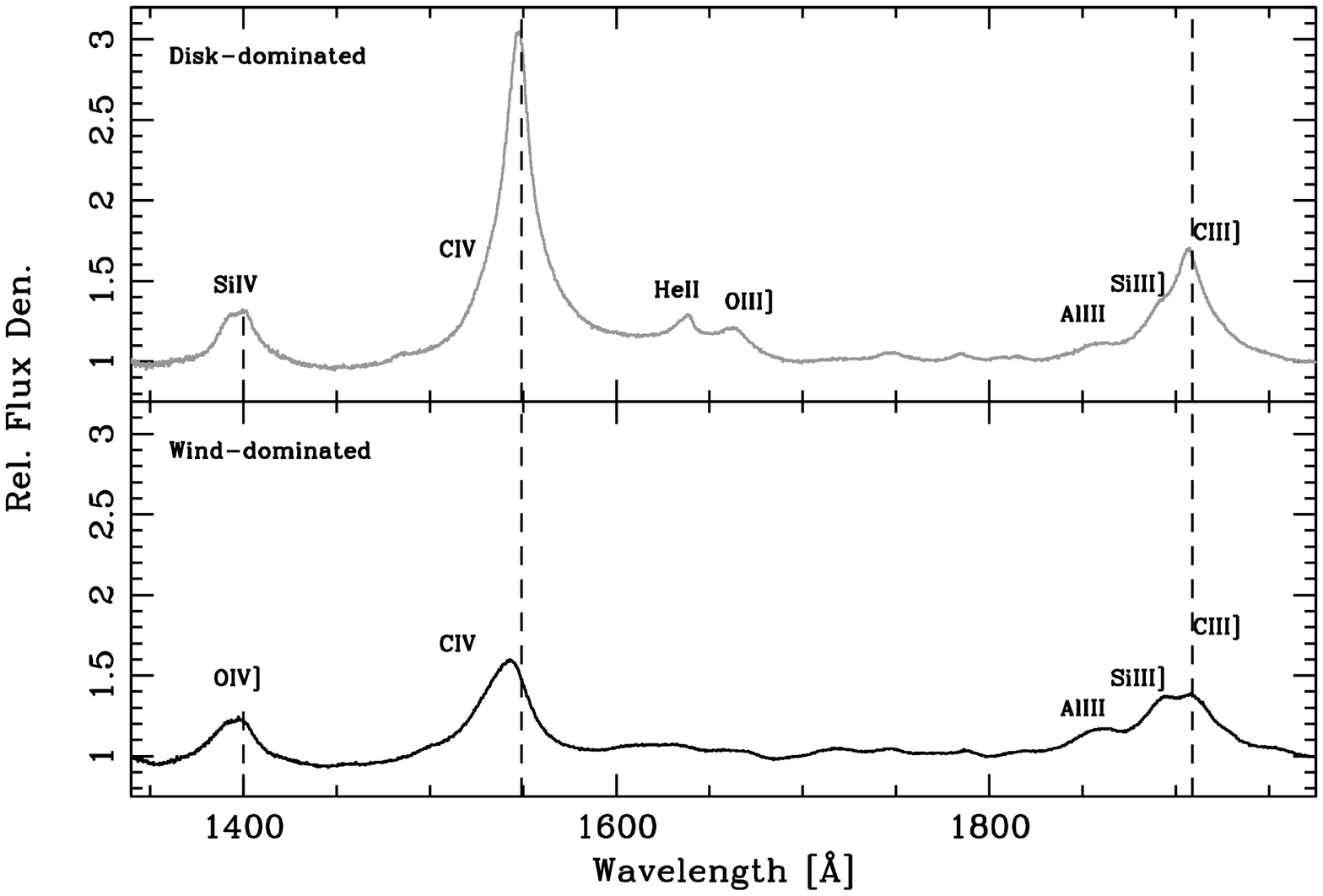}{fig:fig2}{Composite spectra illustrating the extrema of UV emission lines in ``disk-dominated'' ({\em top}) and ``wind-dominated'' ({\em bottom}) quasars.  The disk-dominated objects have strong SiIV, CIV, and CIII], with relatively weak AlIII].  The very strong HeII emission is a telltale signature of a significant contribution to the continuum at $\sim50\,{\rm eV}$.  In the wind-dominated objects, blueshifted OIV] replaces SiIV \citep{lm04}, obscuring the Baldwin Effect in that line.  CIV is both weaker and blueshifted.  The relative lack of HeII indicates a relative lack of $\sim50\,{\rm eV}$ flux as compared to the disk-dominated objects.  Finally, CIII] is weaker, while AlIII has increased in strength, which is expected if this line complex acts as an SED indicator \citep{clb06}.  Overall, the changes in the line features between these extrema are broadly consistent with the notion that the disk-dominated sources are ``hard-spectrum'' and the wind-dominates sources are ``soft-spectrum''; much of the difference may be in the unseen EUV continuum \citep{nd79,kfb97}.}

\section{Implications for Bolometric Luminosities and Black Hole Masses}

This model suggests that there is a range of SEDs for luminous, type 1 quasars and not a single, universal SED.  
The lack of a universal SED has obvious implications for correcting monochromatic luminosities to bolometric luminosities.  Instead of simply considering a universal SED, it is important to chose an SED that is consistent with the line emission from each object.  We suggest that this will cause traditional bolometric corrections to be underestimated for objects that we classify as ``hard-spectrum'' sources and overestimated for objects that we classify as ``soft-spectrum'' sources (Krawczyk et al.\ 2012, in preparation).    These adjustments to the bolometric correction will be small on average, but are systematic and could be important when investigating the differences in Eddington ratios between extreme objects (e.g., RL quasars vs. BALQSOs).  

In terms of black hole masses, this work introduces two areas of concern.  First, as shown by \citet{Richards2011}, the small sample of quasars which have detailed reverberation mapping and form the backbone of the scaling relations used to estimate black hole masses for all other quasars are biased towards what we refer to as hard-spectrum, disk-dominated sources.  Thus it is important to determine whether scaling relations derived from those objects are fully applicable to soft-spectrum, wind-dominated objects.  However, a reverberation mapping campaign of such objects may be prohibitive if evolutionary trends make them more scarce in the present day.

Second, the aforementioned coupling between the disk and the wind could be important for the use of luminosity as a surrogate for radius in black hole mass scaling relations.  Namely, two objects with the same monochromatic luminosity, but very different wind structures, may have disk components of the BELR that are different---including different characteristic radii.  Once again, this argues for the importance of reverberation mapping of apparently wind-dominated quasars.

\acknowledgements The work herein was supported by years of conversations with colleagues, but particularly Karen Leighly, Sarah Gallagher, Pat Hall, Daniel Proga, Paul Hewett, Niel Brandt, Yue Shen, Mike Eracleous, Dan Vanden Berk and Nick Kruczek and was funded in part by NASA-ADP and an Alfred P. Sloan Research Fellowship.

\bibliography{gtrichards}

\begin{thebibliography}{}
\expandafter\ifx\csname natexlab\endcsname\relax\def\natexlab#1{#1}\fi
\expandafter\ifx\csname url\endcsname\relax
  \def\url#1{\texttt{#1}}\fi
\expandafter\ifx\csname urlprefix\endcsname\relax\def\urlprefix{URL }\fi
\providecommand{\eprint}[2][]{\url{#2}}

\bibitem[{{Baldwin}(1977)}]{Baldwin77}
{Baldwin}, J.~A. 1977, \apj, 214, 679

\bibitem[{{Boroson} \& {Green}(1992)}]{bg92}
{Boroson}, T.~A., \& {Green}, R.~F. 1992, \apjs, 80, 109

\bibitem[{{Brotherton} \& {Francis}(1999)}]{bf99}
{Brotherton}, M.~S., \& {Francis}, P.~J. 1999, in Quasars and Cosmology, edited
  by {G.~Ferland \& J.~Baldwin}, vol. 162 of ASP Conf. Series, 395

\bibitem[{{Casebeer} et~al.(2006){Casebeer}, {Leighly}, \& {Baron}}]{clb06}
{Casebeer}, D.~A., {Leighly}, K.~M., \& {Baron}, E. 2006, \apj, 637, 157

\bibitem[{{Collin-Souffrin} et~al.(1988){Collin-Souffrin}, {Dyson}, {McDowell},
  \& {Perry}}]{cdm+88}
{Collin-Souffrin}, S., {Dyson}, J.~E., {McDowell}, J.~C., \& {Perry}, J.~J.
  1988, \mnras, 232, 539

\bibitem[{{Gaskell}(1982)}]{gas82}
{Gaskell}, C.~M. 1982, \apj, 263, 79

\bibitem[{{Hewett} \& {Foltz}(2003)}]{hf03}
{Hewett}, P.~C., \& {Foltz}, C.~B. 2003, \aj, 125, 1784

\bibitem[{{Hewett} \& {Wild}(2010)}]{HW10}
{Hewett}, P.~C., \& {Wild}, V. 2010, \mnras, 405, 2302

\bibitem[{{K{\"o}nigl}(2006)}]{kon06}
{K{\"o}nigl}, A. 2006, MmSAI, 77, 598

\bibitem[{{Korista} et~al.(1997){Korista}, {Ferland}, \& {Baldwin}}]{kfb97}
{Korista}, K., {Ferland}, G., \& {Baldwin}, J. 1997, \apj, 487, 555

\bibitem[{{Kruczek} et~al.(2011){Kruczek}, {Richards}, {Gallagher}, {Deo},
  {Hall}, {Hewett}, {Leighly}, {Krawczyk}, \& {Proga}}]{krg+11}
{Kruczek}, N.~E., {Richards}, G.~T., {Gallagher}, S.~C., {Deo}, R.~P., {Hall},
  P.~B., {Hewett}, P.~C., {Leighly}, K.~M., {Krawczyk}, C.~M., \& {Proga}, D.
  2011, \aj, 142, 130

\bibitem[{{Leighly}(2004)}]{Leighly04}
{Leighly}, K.~M. 2004, \apj, 611, 125

\bibitem[{{Leighly} \& {Casebeer}(2007)}]{lc07}
{Leighly}, K.~M., \& {Casebeer}, D. 2007, in The Central Engine of Active
  Galactic Nuclei, edited by {L.~C.~Ho \& J.-W.~Wang}, vol. 373 of ASP Conf.
  Series, 365

\bibitem[{{Leighly} \& {Moore}(2004)}]{lm04}
{Leighly}, K.~M., \& {Moore}, J.~R. 2004, \apj, 611, 107

\bibitem[{{Murray} et~al.(1995){Murray}, {Chiang}, {Grossman}, \&
  {Voit}}]{mcgv95}
{Murray}, N., {Chiang}, J., {Grossman}, S.~A., \& {Voit}, G.~M. 1995, \apj,
  451, 498

\bibitem[{{Netzer} \& {Davidson}(1979)}]{nd79}
{Netzer}, H., \& {Davidson}, K. 1979, \mnras, 187, 871

\bibitem[{{Proga}(2003)}]{Proga03}
{Proga}, D. 2003, \apj, 585, 406

\bibitem[{{Proga} et~al.(2000){Proga}, {Stone}, \& {Kallman}}]{psk00}
{Proga}, D., {Stone}, J.~M., \& {Kallman}, T.~R. 2000, \apj, 543, 686

\bibitem[{{Reichard} et~al.(2003){Reichard}, {Richards}, {Hall}, {Schneider},
  {Vanden Berk}, {Fan}, {York}, {Knapp}, \& {Brinkmann}}]{rrh+03}
{Reichard}, T.~A., {Richards}, G.~T., {Hall}, P.~B., {Schneider}, D.~P.,
  {Vanden Berk}, D.~E., {Fan}, X., {York}, D.~G., {Knapp}, G.~R., \&
  {Brinkmann}, J. 2003, \aj, 126, 2594

\bibitem[{{Richards}(2006)}]{Richards2006}
{Richards}, G.~T. 2006, ArXiv Astrophysics e-prints.
  \eprint{arXiv:astro-ph/0603827}

\bibitem[{Richards et~al.(2002)Richards, Berk, Reichard, Hall, Schneider,
  SubbaRao, Thakar, \& York}]{rvr+02}
Richards, G.~T., Berk, D. E.~V., Reichard, T.~A., Hall, P.~B., Schneider,
  D.~P., SubbaRao, M., Thakar, A.~R., \& York, D.~G. 2002, AJ, 124, 1

\bibitem[{{Richards} et~al.(2011){Richards}, {Kruczek}, {Gallagher}, {Hall},
  {Hewett}, {Leighly}, {Deo}, {Kratzer}, \& {Shen}}]{Richards2011}
{Richards}, G.~T., {Kruczek}, N.~E., {Gallagher}, S.~C., {Hall}, P.~B.,
  {Hewett}, P.~C., {Leighly}, K.~M., {Deo}, R.~P., {Kratzer}, R.~M., \& {Shen},
  Y. 2011, \aj, 141, 167

\bibitem[{{Sulentic} et~al.(2007){Sulentic}, {Bachev}, {Marziani}, {Negrete},
  \& {Dultzin}}]{sbm+07}
{Sulentic}, J.~W., {Bachev}, R., {Marziani}, P., {Negrete}, C.~A., \&
  {Dultzin}, D. 2007, \apj, 666, 757

\bibitem[{{Sulentic} et~al.(2000){Sulentic}, {Marziani}, \&
  {Dultzin-Hacyan}}]{smd00}
{Sulentic}, J.~W., {Marziani}, P., \& {Dultzin-Hacyan}, D. 2000, \araa, 38, 521

\end{thebibliography}
\bibliographystyle{asp2010}

\end{document}